\documentclass[useAMS,usenatbib,usegraphicx]{mn2e}
\usepackage{url}
\usepackage{times}
\usepackage[usenames]{color}
\usepackage{amsmath}

\title[RR\,Lyrae stars observed with {\it Kepler}]
{Flavours of variability: 29 RR\,Lyrae stars observed with 
{\itshape {\bfseries Kepler}}}
\author[Benk\H{o} et al.]
{J.~M. Benk\H{o}$^{1}$\thanks{E-mail: benko@konkoly.hu},
K. Kolenberg$^{2}$, R. Szab\'o$^{1}$, D.~W. Kurtz$^{3}$, 
S. Bryson$^{4}$,  J. Bregman$^{4}$, \newauthor
M. Still$^{4}$, R. Smolec$^{2}$, J. Nuspl$^{1}$, 
J. Nemec$^{5}$, P. Moskalik$^{6}$, G. Kopacki$^{7}$, 
Z. Koll\'ath$^{1}$, \newauthor
E. Guggenberger$^{2}$, M. Di~Criscienzo$^{8}$,  
J. Christensen-Dalsgaard$^{9}$, H. Kjeldsen$^{9}$, \newauthor
W.~J. Borucki$^{4}$, 
D. Koch$^{4}$, J.~M. Jenkins$^{10}$, and J.~E. Van~Cleve$^{10}$\\
$^{1}$Konkoly Observatory, H-1525 Budapest, P.O. Box 67, Hungary\\
$^{2}$Institut f\"ur Astronomie, Universit\"at Wien, T\"urkenschanzstrasse 17, A-1180 Vienna, Austria\\ 
$^{3}$Jeremiah Horrocks Institute of Astrophysics, University of Central Lancashire, Preston PR1\,2HE, UK\\
$^{4}$NASA Ames Research Center, MS 244-30, Moffett Field, CA 94035, USA\\
$^{5}$Department of Physics \& Astronomy, Camosun College, Victoria, British Columbia, Canada\\
$^{6}$Copernicus Astronomical Centre, ul. Bartycka 18, 00-716 Warsaw, Poland\\
$^{7}$Instytut Astronomiczny Uniwersytetu Wroc{\l}awskiego, Kopernika 11, 51-622 Wroc{\l}aw, Poland\\
$^{8}$INAF-Osservatorio Astronomico di Roma, via Frascati 33, Monte Porzio Catone, Italy\\
$^{9}$Department of Physics and Astronomy, Aarhus University, DK-8000 Aarhus C, Denmark\\
$^{10}$SETI Institute/NASA Ames Research Center, MS 244-30, Moffett Field, CA 94035, USA}
\begin{document}

\date{Accepted 2010 July 22 Received 2010 July 21; in original form 2010 June 4}

\pagerange{\pageref{firstpage}--\pageref{lastpage}} \pubyear{2010}

\maketitle

\label{firstpage}

\begin{abstract}
We present our analysis of {\it Kepler} observations of 29 RR Lyrae  
stars, based on 138-d of observation.  We report precise pulsation  
periods for all stars.  Nine of these stars had incorrect or unknown
periods in the literature. Fourteen of the stars exhibit both
amplitude and phase Blazhko modulations, with Blazhko periods ranging from 27.7
to more than 200\,days. For V445\,Lyr, a longer secondary 
variation is also observed in addition to its 53.2-d Blazhko period. 
The unprecedented precision of the {\it Kepler} photometry has led 
to the discovery of the the smallest modulations detected so far. Moreover, 
 additional frequencies beyond the well-known harmonics and Blazhko 
multiplets have been found. These frequencies are located around the half-integer 
multiples of the main pulsation frequency for at least three stars. In four stars, these 
frequencies are close to the first and/or second overtone modes. The amplitudes of 
these periodicities seem to vary over the Blazhko cycle. V350\,Lyr, a 
non-Blazhko star in our sample, is the first example of a double mode RR\,Lyrae star that 
pulsates in its fundamental and second overtone modes.
\end{abstract}

\begin{keywords}
stars: oscillations --- stars: variables: RR Lyrae
\end{keywords}

\section{Introduction}

\begin{table*}
\begin{center}
\caption{Basic parameters of 29 RR\,Lyrae stars observed by
 {\it Kepler} (long--cadence) during the runs from Q0 to Q2. \label{All_stars}}
\begin{tabular}{@{}r*{4}{@{\hspace{7pt}}c}
@{\hspace{5pt}}l@{\hspace{5pt}}r@{\hspace{5pt}}l@{}*{3}{@{\hspace{5pt}}l}@{}}
\hline
KIC &  RA        &   DEC    & $K_{\mathrm p}$   & $P_0$  &  $\sigma(P_0)$ & $A_1$ &  $\sigma(A_1)$  & 
\hspace{0.5cm} ID  &  Runs &  Note  \\ 
    &  [J2000]   &  [J2000] &  [mag]  &  [d] & [$10^{-5}$\,d]  &   [mag]   & [$10^{-3}$\,mag] & & & \\ 
\hline
3733346 & 19 08 27.23 &    +38 48 46.19 &   12.684 & 0.68204 & 1.68&0.266 & 2.2&  NR~Lyr    &  Q1,Q2 &     \\
3864443 & 19 40 06.96 &    +38 58 20.35 &   15.593 & 0.48680 & 0.73&0.331 & 2.4&  V2178~Cyg & Q1,Q2  & Bl,n \\
3866709 & 19 42 08.00 &    +38 54 42.30 &   16.265 & 0.47071 & 0.85&0.340 & 3.0&  V715~Cyg  &  Q1,Q2 &    \\
4484128 & 19 45 39.02 &    +39 30 53.42 &   15.363 & 0.54787 & 1.24&0.299 & 2.9&  V808~Cyg  & Q1,Q2 & Bl \\
5299596 & 19 51 17.00 &    +40 26 45.20 &   15.392 & 0.52364 & 0.91&0.195 & 1.5&  V782~Cyg  & Q1,Q2  &     \\
5559631 & 19 52 52.74 &    +40 47 35.45 &   14.643 & 0.62072 & 1.48&0.273 & 2.4&  V783~Cyg  & Q1,Q2  &  Bl \\
6070714 & 19 56 22.91 &    +41 20 23.53 &   15.370 & 0.53410 & 0.93&0.229 & 1.7&  V784~Cyg  & Q1,Q2  &     \\
6100702 & 18 50 37.73 &    +41 25 25.72 &   13.458 & 0.48815 & 0.71&0.206 & 1.5&  ASAS~185038+4125.4  & Q0,Q1,Q2 & \\
6183128 & 18 52 50.36 &    +41 33 49.46 &   16.260 & 0.56168 & 1.07&0.245 & 1.9&  V354~Lyr  & Q1,Q2 & Bl,n \\
6186029 & 18 58 25.69 &    +41 35 49.45 &   17.401 & 0.51293 & 1.13&0.204 & 2.0&  V445~Lyr  & Q1,Q2 & Bl,n \\
6763132 & 19 07 48.37 &    +42 17 54.67 &   13.075 & 0.58779 & 1.13&0.280 & 2.3&  NQ~Lyr    & Q0,Q1,Q2 &  \\
7030715 & 19 23 24.53 &    +42 31 42.35 &   13.452 & 0.68362 & 1.43&0.231 & 1.8&  ASAS~192325+4231.7   & Q0,Q1,Q2  & \\
7176080 & 18 49 24.43 &    +42 44 45.56 &   17.433 & 0.50708 & 0.94&0.357 & 3.0&  V349~Lyr             & Q1,Q2     & Bl,n \\
7198959 & 19 25 27.91 &    +42 47 03.73 &   $\phantom{0}$7.862  & 0.56688 & 1.26&0.239 & 2.2&   RR~Lyr              & Q1,Q2     & Bl \\
7505345 & 18 53 25.90 &    +43 09 16.45 &   14.080 & 0.47370 & 1.29&0.374 & 3.4&   V355~Lyr            & Q2        & Bl \\
7671081 & 19 09 36.63 &    +43 21 49.97 &   16.653 & 0.50457 & 0.86&0.313 & 2.5&   V450~Lyr            & Q1,Q2     & Bl \\
7742534 & 19 10 53.40 &    +43 24 54.94 &   16.002 & 0.45649 & 0.77&0.407 & 3.5&   V368~Lyr            & Q1,Q2     & n \\
7988343 & 19 59 50.67 &    +43 42 15.52 &   14.494 & 0.58115 & 1.28&0.341 & 3.0&   V1510~Cyg           & Q1,Q2     &  \\
8344381 & 18 46 08.64 &    +44 23 13.99 &   16.421 & 0.57683 & 1.26&0.322 & 2.9&   V346~Lyr            & Q1,Q2     &  \\
9001926 & 18 52 01.80 &    +45 18 31.61 &   16.914 & 0.55682 & 1.13&0.287 & 2.5&   V353~Lyr            & Q1,Q2     & Bl,n \\
9508655 & 18 49 08.37 &    +46 11 54.96 &   15.696 & 0.59424 & 1.33&0.339 & 3.0&   V350~Lyr            & Q1,Q2     &    \\
9578833 & 19 09 40.64 &    +46 17 18.17 &   16.537 & 0.52702 & 0.99&0.304 & 2.5&   V366~Lyr            & Q1,Q2     & Bl,n \\
9591503 & 19 33 00.91 &    +46 14 22.85 &   13.293 & 0.57139 & 1.09&0.384 & 3.2&   V894~Cyg            & Q0,Q1,Q2  &  \\
9697825 & 19 01 58.63 &    +46 26 45.74 &   16.265 & 0.55759 & 1.06&0.261 & 2.1&   V360~Lyr            & Q1,Q2     &   Bl,n \\
9947026 & 19 19 57.96 &    +46 53 21.41 &   13.300 & 0.54859 & 0.88&0.219 & 1.6&   V2470~Cyg           & Q0,Q1,Q2  &   \\
10136240 & 19 19 45.28 &   +47 06 04.43 &   15.648 & 0.56579 & 1.23&0.270 & 2.4&   V1107~Cyg           & Q1,Q2     & n  \\
10789273 & 19 14 03.90 &   +48 11 58.60 &   13.770 & 0.48029 & 0.86&0.390 & 3.4&   V838~Cyg            & Q1,Q2  & \\
11125706 & 19 00 58.77 &   +48 44 42.29 &   11.367 & 0.61323 & 0.98&0.179 & 1.2&   ROTSE1~J190058.77+484441.5 & Q0,Q1,Q2 & Bl \\
12155928 & 19 18 00.49 &   +50 45 17.93 &   15.033 & 0.43639 & 0.71&0.394 & 3.4&   V1104~Cyg           & Q1,Q2      & Bl \\
\hline                                                   
\end{tabular}
\end{center}                                            
\end{table*}

The ultraprecise photometry by the {\it Kepler} space telescope
opens up the possibility of discovering new phenomena and shedding new light on 
long-standing astrophysical problems \citep{Gil10}. One of the most interesting unsolved problems is 
the physical origin of the \citet{Bla07} effect, an amplitude and/or phase modulation 
of RR\,Lyrae stars. The leading explanations are: (1) an oblique rotator model 
that invokes a magnetic field  \citep{Shi00}; 
(2) a model with resonant coupling between radial and non-radial 
mode(s) \citep{DzM04}; and (3) a mechanism invoking a cyclic variation of the 
turbulent convection caused by a transient magnetic field \citep{Sto06}.
At present none of these models explains all the observed properties of 
stars showing the Blazhko effect. It is not even clear whether a modification of 
the above ideas or new astrophysical processes are needed to solve the problem. 
Comprehensive discussions of the observational and theoretical properties 
of Blazhko RR\,Lyrae stars are given by \cite{Sze88} and \citet{Kov09}.

This paper paper describes new properties of RR\,Lyrae stars revealed by  
early data from the {\it Kepler} photometer.
Here we concentrate on the results based mostly on Fourier analyses. 
A detailed study of all observed stars is beyond the scope of the present paper.

\section{Data}

A detailed technical description of the {\it Kepler Mission} can be found in 
\cite{Koch10} and \cite{Jen10a, Jen10b}. At the time of this writing three long 
cadence (29.4-min integration time) photometric data sets have been released to 
the KASC (Kepler Asteroseismic Science Consortium). Altogether 29 RR\,Lyrae 
stars were observed in this way. One additional RR\,Lyrae star 
(V355\,Lyr) has been released since the publication by \citet{Kol10}.
 This source was observed as a Director's discretionary
target\footnote{http://keplergo.arc.nasa.gov/GOprogramDDT.shtml} in
the {\it Kepler} Guest Observer Program\ from quarter 2 onwards, and is
included here to enrich the KASC sample. Throughout the
work presented here, we used only the long--cadence data\footnote{We have observed 
several additional RR\,Lyrae candidates with short cadence (1-min integration);
results from these data will be discussed in a forthcoming paper.}.

The commissioning phase data (Q0) included six of the RR\,Lyrae stars between 2009 
May 2 to 11  (9.7\,d), the observations of Q1 data began on 2009 May 13 and ended 
on 2009 June 15 (33.5\,d). The first full quarter of data (Q2) ran from 
2009 June 19 to 2009 September 16 (89\,d). At present, the combined number of data 
points for a given star is between 4096 and 6175. Column 8 in Table~\ref{All_stars} 
shows the available data sets for each star\footnote{Public {\it Kepler} data can be downloaded
from the web page: \\ http://archive.stsci.edu/kepler/}.

\begin{figure*}
\includegraphics[width=16cm]{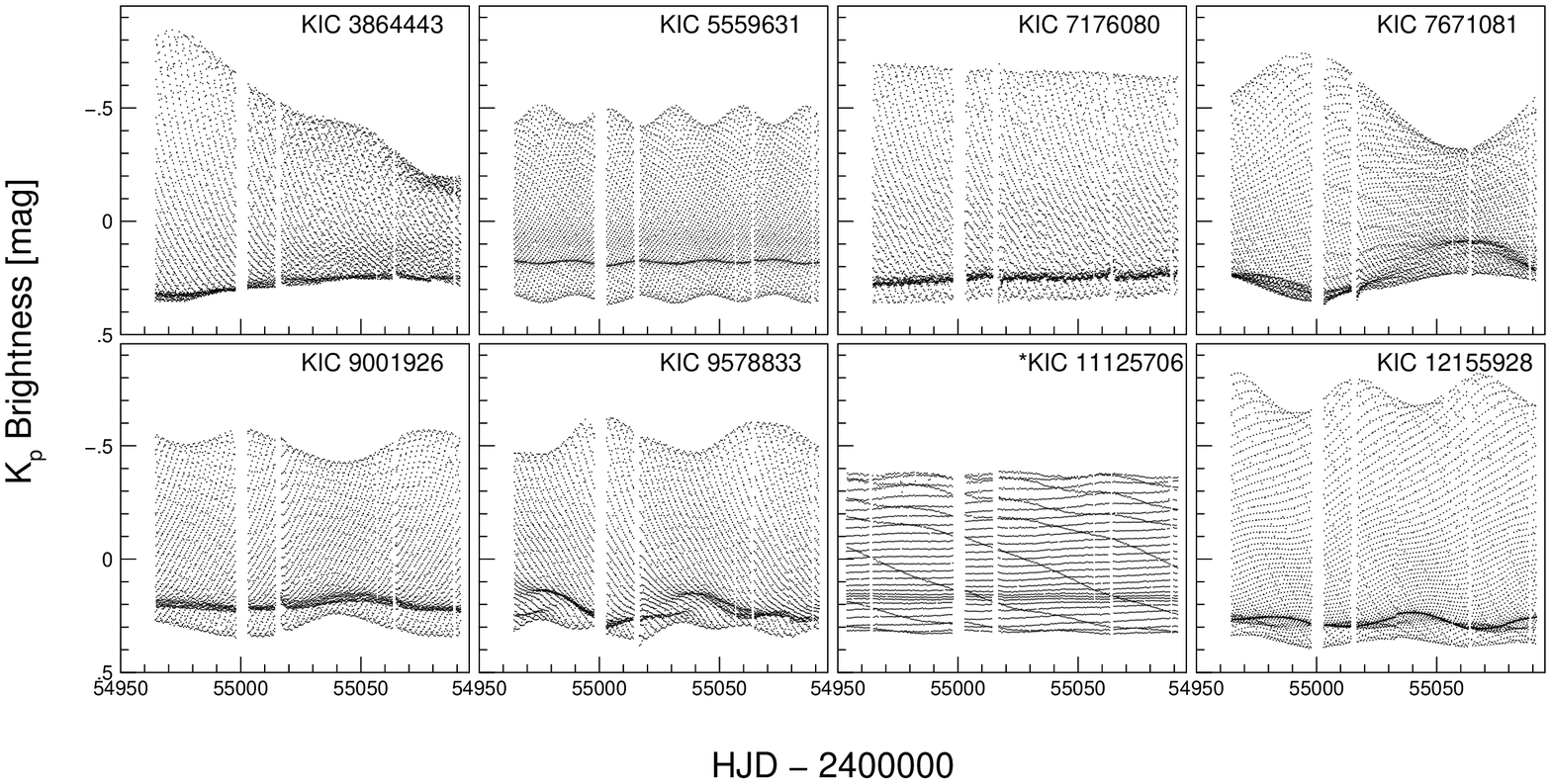}
\caption[]{
The gallery of {\it Kepler} Blazhko stars. The figure shows the complete 
light curves of eight stars observed with long--cadence during the periods Q0 
through Q2. Further light curves are given in Fig.\,\ref{double} and 
Fig.\,\ref{add} and in our parallel papers \citep{Sza10,Kol10b}. 
Lines seen running through these light curves are visual 
artefacts caused by beating of the sampling frequency with the pulsation 
frequencies. They cause no problems in the Fourier analysis. $^{*}$For better 
visibility the scale of KIC\,11125706 is increased by a factor of 1.5.
} \label{zoo}
\end{figure*}

\begin{table*}
\begin{center}
\caption{Period, amplitude and phase properties of the Blazhko stars}
\label{Blazhko_stars}
\begin{tabular}{llr@{\hspace{5pt}}crrrl@{}}
\hline
KIC & GCVS  &    P$_{\mathrm B}$  & $\sigma(P_{\mathrm B})$&  
$\Delta A_1$   &  $\Delta \phi_1$  & Q & Addition. freq.$^{(b)}$ \\ 
    &       &    [d]    & [d]  & [mag]   &     &  &   \\ 
\hline
3864443 & V2178\,Cyg &  $>200$	&   & $>0.488$   & $>0.0014$ &  0.153 & F2, (PD) \\
4484128 & V808\,Cyg  &  $\approx90$  & &  0.304       & 0.012142  & $-0.045$ & PD \\
5559631 & V783\,Cyg  &  27.7	    & 0.4 & 	0.071  &	0.001591 & 0.156 &  \\
6183128 & V354\,Lyr  &  $\gg 127$&   & $>0.245$&	$>0.00245$& $-0.139$  & F2, (F1, PD, F') \\
6186029 & V445\,Lyr  &  53.2$^{(a)}$ & 2.8 & 0.968& 0.022442 & 0.540 & PD, F1, F2 \\ 
7176080 & V349\,Lyr  & $\gg 127 $& & $>0.060$ & $>0.00175$ & $-0.251$ &  \\
7198959 & RR\,Lyr    & 39.6      & 1.8 &  0.461   & 0.013836 & 0.676 &  PD \\
7505345 & V355\,Lyr  & 31.4      & 0.1 &  0.107   & 0.003518 & 0.136 &  PD \\
7671081 & V450\,Lyr  & $\approx125$& & 0.391   & 0.004882 & 0.235 &  \\
9001926 & V353\,Lyr  & 60.0  & 7.1 &  0.157     & 0.004026 &0.033 &  \\
9578833 & V366\,Lyr  & 65.6  & 2.6 &  0.171    & 0.003205 & $-0.162$ & \\
9697825 & V360\,Lyr  & 51.4  & 4.3 &    0.356    & 0.008228 & 0.279  & F1, (PD) \\
11125706 &  	     & 39.4  & 2.0 &  0.030      & 0.001420 & 0.540 & \\
12155928 & V1104\,Cyg  & 53.1 & 0.3 &  0.105     & 0.002451 & $-0.063$ & \\
\hline                                                   
\end{tabular}

\footnotesize{
$^{(a)}$The star shows a longer time-scale variation than its Blazhko modulation as well.

$^{(b)}$The pattern of additional frequencies: PD means period doubling; 
F1 indicates first overtone frequency and its linear combination with
fundamental one; F2 is as F1, but 
with second overtone; F$^\prime$ indicates frequencies with unidentified modes; 
brackets indicate marginal effects.
}
\end{center}                                            
\end{table*} 

The telescope is rotated by 90 degrees four times per orbital period for best 
exposure of the solar panels. Accordingly, the Q0 and Q1 data sets were observed 
at one position and the first roll was executed between Q1 and Q2. The different 
apertures applied to a given star in the two positions caused the zero points of 
the raw fluxes and the amplitudes of a star to be different for the two rolls. We 
applied simple linear transformations to fit the amplitudes and zero points for 
combining the data. The long time-scale trends were removed from the raw fluxes by 
a trend filtering algorithm prepared for CoRoT RR\,Lyrae 
data \citep{Cha10}\footnote{http://www.konkoly.hu/HAG/Science/index.html}, then 
fluxes were transformed into a magnitude scale, where the averaged magnitude 
of each star was fixed to zero. Measured raw fluxes are in the range of 
$3.1\times 10^{10}>F>1.3\times 10^6$\,ADU which yields 
$6\times 10^{-6}$--$9\times 10^{-4}$\,mag accuracy for an individual data point.

\section{Analysis and results}

As a first step Fourier analyses were performed on the data sets. To this end we 
used the software packages {\sc MuFrAn} \citep{Kol90}, {\sc Period04} \citep{LB05} and 
{\sc SigSpec} \citep{Re07}, all of which gave similar frequency spectra with similar errors. 
Some basic parameters of the observed stars are summarized in 
Table~\ref{All_stars}. The columns of the table shows the ID numbers, J2000 positions 
(RA, DEC), and apparent magnitude ({\it Kepler} magnitude $K_{\mathrm p}$), 
all from the Kepler Input 
Catalog (KIC-10)\footnote{http://archive.stsci.edu/kepler/kic10/search.php}.
The next columns contain the main pulsation periods and the Fourier 
amplitude of the main frequencies obtained from our {\sc SigSpec} analysis. 
These two basic parameters ($P_0$ and $A_1$) were previously unknown or wrong for nine stars 
(signed by an `n' in the last column). 

This fact already suggests, that there were hardly any detailed investigations
on these stars before this work. This is underlined by the lack of known stars
showing Blazhko effect before {\it Kepler} except RR\,Lyr itself. In most of
the cases these stars are mentioned in the literature apropos of their discoveries
and in some cases due to the determination of their position and/or ephemeris.
The radial velocity measurements of NR\,Lyr and V894\,Cyg were used for
kinematic study of the Galaxy \citep{Lay94, Bee00, Jef07}. 
The Northern Sky Variablity Survey (NSVS) included NR\,Lyr, V355\,Lyr 
and V2470\,Cyg. Their NSVS light curves -- together with more than 1100 other 
ones -- formed the basis of the statistical study of \cite{Kin06}. 
The most specific investigations were carried out on V783\,Cyg by \cite{Los79} 
and \cite{Cro91}, who found a period increase between 1933 and 1990 with the ephemeris of
${\mathrm JD_{max}}=2436394.332 + 0.62069669 E + 7.5 \cdot 10^{-11} E^2$.
According to Table~\ref{All_stars} the period is still increasing with a
good agreement of the rate of Cross's value: $0.088\pm0.023$~d My$^-1$.

The standard errors of the main period and amplitude (in columns 6 and 8) were 
estimated from the accuracy of non-linear least-square fits. The last three columns 
of the table indicate other identifications of the stars, the observing runs analysed 
and the presence of a Blazhko effect, respectively.
All the periods, amplitudes and light curve shapes of the 29 stars are
typical for RRab stars pulsating in their radial fundamental
mode, therefore, these classifications are omitted from Table~\ref{All_stars}.

\begin{figure*} 
\includegraphics[width=16cm]{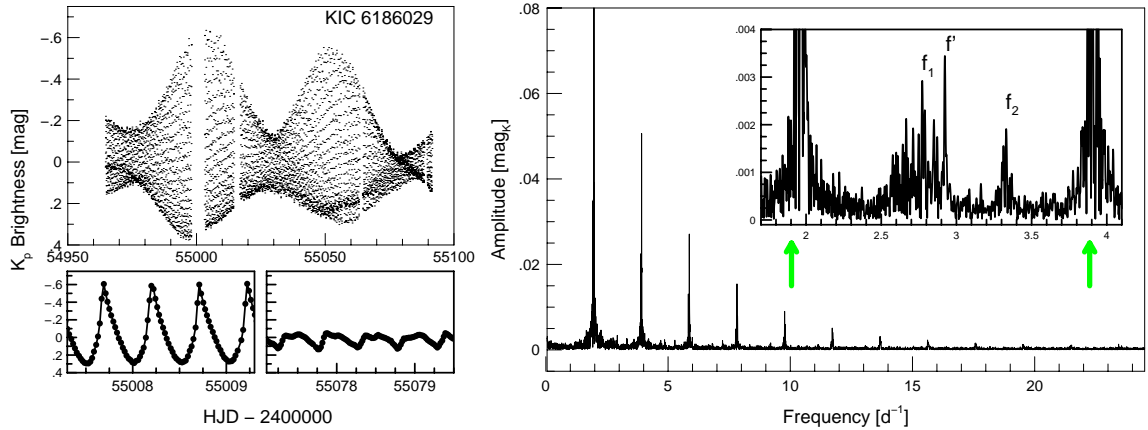}
\caption{{\bf Top left:} Light curve of V445\,Lyr (KIC\,6186029). {\bf Bottom 
left:} Parts of the light curve from maximum and minimum of the amplitude of the 
Blazhko cycle. {\bf Right:} Fourier spectrum after the data are prewhitened with the main 
frequency and its harmonics. The insert is a zoom around the positions of the main 
frequency $f_0$ and its first harmonics $2f_0$ (showed by green arrows) after the 20 
highest amplitude frequencies were removed. 
The complex structure of additional frequencies is clearly seen.
}\label{double}
\end{figure*}

\subsection{RR\,Lyrae stars with amplitude modulation}

Generally, it is an easy task to distinguish amplitude modulated and non-modulated 
{\it Kepler} RR\,Lyrae light curves. A gallery of modulated light curves is shown 
in Fig.\,\ref{zoo}. It is obvious at first glance that the modulation cycles are 
predominantly long and the amplitude of the effect is clearly visible. 
Non-sinusoidal envelopes of the light curves (see, e.g., V2178\,Cyg; KIC\,3864443) or 
moving bumps (e.g., V366\,Lyr; KIC\,9578833) are also conspicuous.

\begin{figure}
\includegraphics[width=8cm]{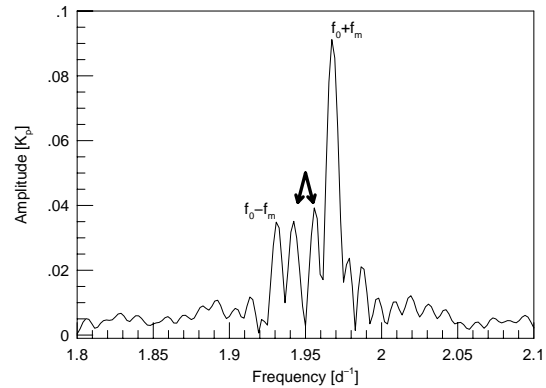}
\caption[]{
Fourier amplitude spectrum of V445\,Lyr (KIC\,6186029) around 
the main pulsation frequency $f_0$ after the data are prewhitened with the
main frequency. Beyond elements of the Blazhko triplet $f_0\pm f_{\mathrm B}$
two additional peaks are seen (arrows). 
} \label{d_sp}
\end{figure}

The interesting light curve of V445\,Lyr (KIC\,6186029) is shown separately in 
Fig.\,\ref{double}. The two observed Blazhko cycles are surprisingly different. 
The high amplitude of the Blazhko modulation extremely distorts the shape of the 
light curve. This is demonstrated in the small panels of Fig.\,\ref{double}. 
Signs of complex variations are detectable from the Fourier spectrum as well. The 
spectrum of the data prewhitened with the main pulsation frequency and its harmonics shows 
four peaks around each of the harmonics (Fig.\,\ref{d_sp}). 
Two outer peaks at the harmonics can be 
identified as elements of the Blazhko triplets ($f_0\pm f_{\mathrm B}$). Two other side peaks 
closer to the harmonic frequencies show the possible variation of the Blazhko 
effect on a time-scale longer than the observation run. This may be a result of a 
cyclic variation (existence of more than one Blazhko modulation), a secular trend, 
or random changes. Several papers have reported multiperiodic and/or unstable 
behaviours in the Blazhko effect (e.g., \citealt{Lac04, Col06, Kol06, NK06, Sod06, Szcz07, Wi08, 
Jur09}). After investigating the two Blazhko cycles noted here, we are not in the 
position to decide on the nature of this variation, but the 3.5--5 year long time 
base of {\it Kepler}'s observations will provide an excellent opportunity to study 
this strange behaviour.

We were systematically searching for low amplitude Blazhko RR\,Lyrae stars. 
Instrumental trends of the observed fluxes that are not properly removed could 
cause apparent amplitude changes in the non-linear magnitude scale.
A decreasing trend of the averaged fluxes results in increasing amplitudes 
in magnitudes and vice-versa. Therefore, we always checked the amplitude variation using 
the raw fluxes. We divided the data sets into small sections (typically 2--3\,d 
in length), then calculated the amplitude difference $\delta A_1(t)$
of the first Fourier component and its averaged value $\Delta A_1$ over the whole time 
span for all sections by a non-linear fit. These calculated functions reflect well the 
variation of pulsation amplitude seen in the light curves.

With the help of this tool we found in KIC\,11125706 the lowest 
amplitude modulation ever detected in an RR\,Lyrae star. Full amplitude of the 
maximum light variation $A(K_{\mathrm p})_{\mathrm max}=0.015$\,mag, and the 
amplitude of the highest side peak in the Fourier spectrum is only 
$A_{K_{\mathrm p}}(f_0+f_{\mathrm B})=0.0022$\,mag. 
The lowest published amplitude of a Blazhko effect 
previously found was the case of DM\,Cyg \citep{JH09} where 
$A(V)_{\mathrm max}=0.07$\,mag, $A_V(f_0+f_{\mathrm B})=0.0096$\,mag 
and $A_I(f_0+f_{\mathrm B})=0.0061$\,mag. 
The two measurements are not strictly comparable because {\it Kepler} passband is 
broad in white light. However, the maximum of its spectral response function 
(see \citealt{Koch10}) is about 6000\,{\AA} between Johnson--Cousins filters $V$ and $R$.

All the Blazhko RR\,Lyrae stars found in our sample are listed in 
Table~\ref{Blazhko_stars}. The estimated lengths of the Blazhko cycles are 
indicated in the third column. For the shorter periods 
than the total time span they were calculated from the 
averaged frequency differences of the highest side peaks 
($f_0+f_{\mathrm B}$ and $f_0-f_{\mathrm B}$), otherwise a minimum period is given. 
The fourth column shows the amplitude 
modulation parameter $\Delta A_1$ defined above.

We note here, the triplet structure has always appeared for all Blazhko
stars, i.e. no stars show frequency doublets. The amplitude of the high frequency 
peak is higher than the lower for 9 stars. The remaining 5 show the opposite pattern.
The asymmetry parameter $Q$ defined by \cite{Al03} 
as $Q=[A(f_0+f_{\mathrm B})-A(f_0-f_{\mathrm B})] [A(f_0+f_{\mathrm 
B})+A(f_0-f_{\mathrm B})]^{-1}$
varies from $-0.251$ to 0.676 (see the 6th column in Table\,\ref{Blazhko_stars}), 
however, these values are rather preliminary due to the long Blazhko cycles.

In the past few years, thanks to high precision ground- and space-based
observations, the known occurrence rate of the Blazhko effect among RR\,Lyrae 
stars has increased from the former estimate of 15--30\% to a ratio close to 50\% 
(see \citealt{Jur09, Cha09, Kol10}). It is even possible that all RR\,Lyrae stars 
may show a Blazhko modulation with an increasing frequency of Blazhko stars at 
lower modulation amplitude. The {\it Kepler} measurements provide an ideal tool to test this 
hypothesis. From our $\Delta A_1$ values in Table~\ref{Blazhko_stars} it can be 
seen that we found only two stars with modulation amplitude lower than $0.1$\,mag.

To test our detection limit of the amplitude modulation, artificial light curves were 
generated. Two sets of grids were constructed: one for V368\,Lyr (KIC\,7742534) 
and another for KIC\,7030715. These stars have the shortest and the longest pulsation periods 
(0.45649\,d and 0.68204\,d) in the sample, respectively. In both cases the Fourier 
parameters of the main pulsation frequency and its significant harmonics were used 
to build the artificial light curves. These were modulated by a simple sinusoidal 
function with amplitudes ranging between $0.1$ and $0.001$\,mag, and with modulation 
periods from 25 to 150\,d with a step size of 25\,d, according to the general 
modulation formula (eq.\,2) in \cite{Ben09}. Measured averaged fluxes of the 
non-Blazhko stars are in the range of $1.8\times 10^8>F>2.7\times 10^6$\,ADU which 
means a noise between $8\times 10^{-5}$ and $6\times 10^{-4}$\,mag. This was taken 
into account by adding Gaussian noise with $\sigma = 10^{-4}$ and 
$5\times 10^{-4}$ to the artificial data. The light curves were always calculated at the 
observed points of time.

In our tests, we reckoned the amplitude modulation as detectable if the highest 
Fourier side peak connected to the modulation exceeds the spectral significance
($\sigma_{\mathrm s}$) 5.
(For the definition of the $\sigma_{\mathrm s}$ we refer \cite{Re07};
the correspondence between more popular amplitude
signal-to-noise ratio $S/N$ \citep{Bre03} and $\sigma_{\mathrm s}$ 
is yielded as $\sigma_{\mathrm s}=5\approx \ S/N=3.83$). 
The obtained limiting values are $A(f_0+f_{\mathrm B}) > 0.001$--$0.002$\,mag 
(or $\Delta A_1 > 0.005$--$0.01$\,mag) depending on the 
brightness, but highly independent of the periods (P$_0$ and P$_{\mathrm B}$). 
Higher sampling rate (i.e. short cadence) does not decrease
our detection limit because the present sampling frequency 48.98\,d$^{-1}$ is 
much higher than a typical Blazhko frequency (0.1--0.01\,d$^{-1}$), hence each 
Blazhko cycle is covered sufficiently. 

\begin{figure}
\includegraphics[width=8cm]{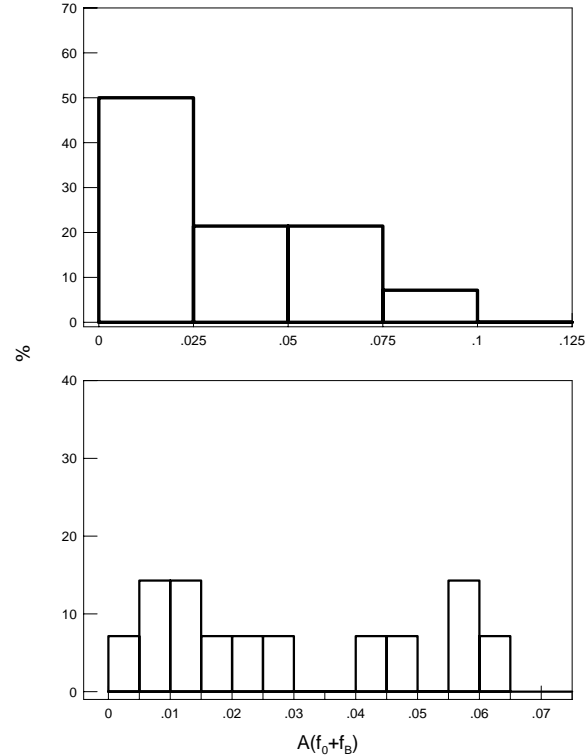}
\caption[]{
Modulation amplitude distribution of the Blazhko variables in 
the {\it Kepler} sample using $0.025$\,mag (top) and $0.005$\,mag 
(bottom) size of bins. The modulation amplitude corresponds to the Fourier 
amplitude of the largest amplitude modulation frequency component $A(f_0+f_{\mathrm B})$.
} \label{hist}
\end{figure}

Notwithstanding our efforts, we did 
not detect any modulation for 15 RR\,Lyrae stars in our {\it Kepler} sample,
however, some very small amplitude modulation with long period might remain undetected. 

Using the same number (14) of Blazhko stars \cite{Jur09} found an exponential-like 
distribution of their modulation amplitude strengths. Our sample 
shows a similar behaviour (top in Fig.\,\ref{hist}) when we divide it up into the 
same $0.025$\,mag size of bins as were used by \cite{Jur09}. However, the 
distribution seems be to more uniform, when we split our sample into smaller size of bins 
(bottom in Fig.\,\ref{hist}). Although, our sample is affected by small number statistics 
in this respect, we checked the uniformity of the modulation amplitude distribution. 
We carried out a one-sample Kolmogorov--Smirnov (K-S) test  
over the intervals of (0, $\Delta A_1^{\mathrm max}$) 
and (0, $A(f_0+f_{\mathrm B})^{\mathrm max}$), respectively.
(The extremely modulated star V445\,Lyr was omitted from the sample.)
Both tests allowed the uniform distribution hypothesis with 99 per-cents of probabilities. 
We note that the K-S test uses data directly, without any binning.

\subsection{Phase modulation}

\begin{figure*}
\includegraphics[width=16cm]{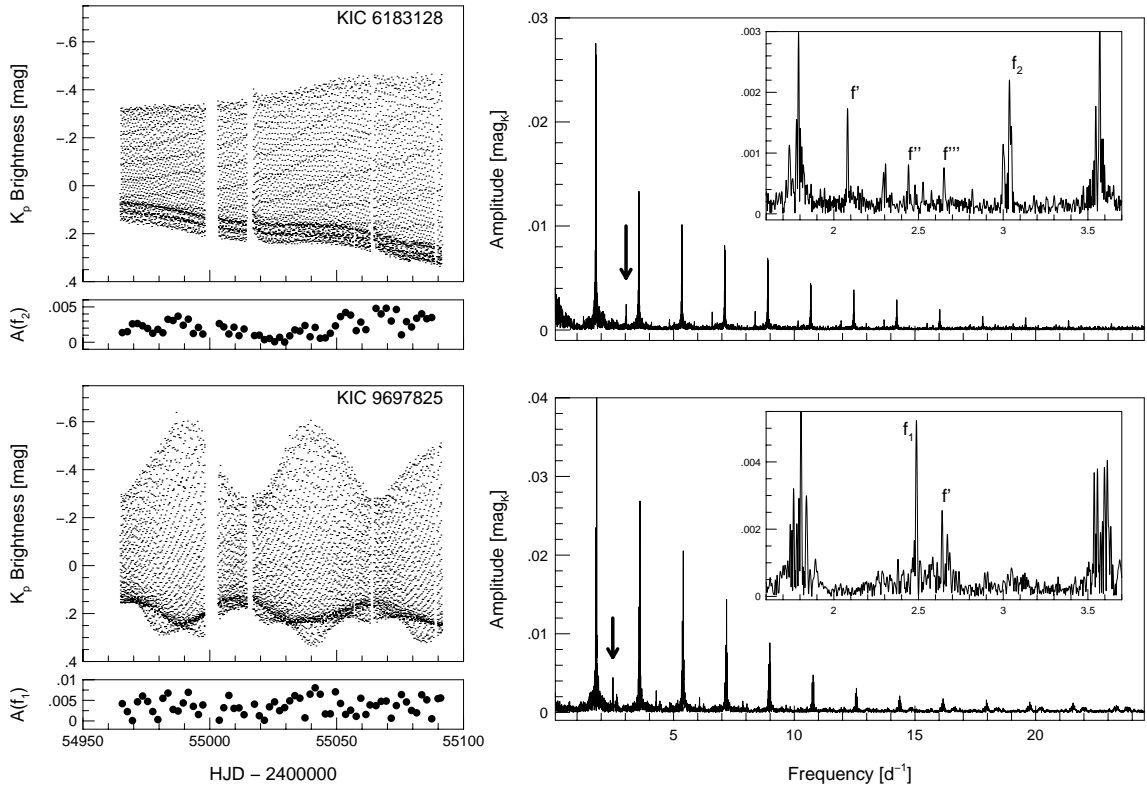}
\caption{{\bf Left top:} Light curves of two {\it Kepler} RR\,Lyrae stars, 
V354\,Lyr (KIC\,6183128) and V360\,Lyr (KIC\,9697825). {\bf Right:} Fourier 
spectra after the data are prewhitened with the main frequency and its harmonics. Arrows point 
to the highest peaks connected to additional frequencies. The inserts are a zoom 
around the position of the highest additional peaks. {\bf Left bottom:} Amplitude 
of the highest additional peaks vs. time.
The typical error is about $0.001$\,mag -- smaller than the size of the data 
points.
}  
\label{add}
\end{figure*}

Without any ``a priori'' knowledge about the nature of the modulation, frequency modulation  
(period changes over the Blazhko cycle) and phase modulation can not be distinguished:
a detected phase variation indicates period changes, and vice-versa.
From now on we refer to this phenomenon as phase modulation.

We searched for phase modulation in all of our RR\,Lyrae stars in the same way as 
was described in the case of amplitude modulation. 
We calculated the $\Delta \varphi_1$ values from the phase variation 
function $\delta\varphi_1(t)$. We expressed the 
phase differences relative to the total cycle, that is 
$\Delta \phi_1$=$\Delta \varphi_1/ 2 \pi$ (= $\delta P_0/ P_0$). 
The results can be seen in the 5th column of Table~\ref{Blazhko_stars}.

We detected clear phase modulation for all the studied Blazhko RR\,Lyrae stars. 
The hardest task was to find it in the case of V2178\,Cyg, where the Blazhko cycle 
is longer than the data set and the phase variation during the observed time span 
was only 0.0014 ($\approx$1 min). The reality of this small phase variation was 
checked and confirmed by the sensitive analytical function method \citep{Kol02}. 
We can detect period variations smaller than 1.5\,min for three further stars: 
V783\,Cyg (KIC\,5559631), V349\,Cyg (KIC\,7176080), and KIC\,1125706. 
The other extremity is represented by V445\,Lyr and RR\,Lyr itself with their 
values of 0.0224 (= 16.6\,min) and 0.0138 
(= 11.3\,min), respectively. There is no clear indication for any connections 
between the strengths of the two type of modulations.

As was shown by \cite{SzJ09} and \cite{Ben09}, phase modulation always causes 
multiplet structures of higher order than triplets in the Fourier spectrum around 
the main frequency and its harmonics. These multiplet peaks were most clearly 
detected for V808\,Cyg, RR\,Lyr and V360\,Lyr, the Blazhko stars with the 
strongest phase modulation in our sample.

The precise and almost continuous observation of 14 Blazhko RR\,Lyrae stars 
measured by {\it Kepler} convincingly demonstrates that in all Blazhko stars both 
amplitude and phase variation are present.

\subsection{Additional frequencies}

Besides the modulation components that occur in multiplet structures around the 
main frequency and its harmonics, we found additional frequencies. In the cases of 
V808\,Cyg, V355\,Lyr and RR\,Lyr these frequencies are located around $f_0/2, 
3f_0/2, 5f_0/2,\dots$, where $f_0$ denotes the main pulsation frequency. This very 
interesting period doubling effect has already been discussed briefly in 
\citet{Kol10} for RR\,Lyr itself, and a separate paper \citep{Sza10} is dedicated 
to it. Here we just remark that the presence of these frequencies in the spectra 
seems to be variable in time and connected to particular Blazhko phases. The 
phenomenon can be described in a purely radial framework of pulsation theory.

Similarly, time-dependent phenomena might also be important for the four further stars 
V354\,Lyr, V2178\,Cyg, V360\,Lyr and V445\,Lyr, where we discovered additional 
frequencies with small amplitudes. The light curves of V354\,Lyr and V360\,Lyr 
are plotted in the top left panels of Fig.\,\ref{add}. The last column of 
Table~\ref{Blazhko_stars} indicates the possible identification of the additional 
frequencies.

In the Fourier spectrum of {\bf V354\,Lyr} (KIC\,6183128) the highest ($\sigma_{\mathrm s}= 34.4$) 
additional peak is at $f_2=3.0369\pm 0.0002$\,d$^{-1}$ (see insert in Fig.\,\ref{add}). 
Its ratio to the fundamental frequency ($f_0=1.78037\pm 0.00004$\,d$^{-1}$) is 0.586, 
which is close to the canonical ratio of the fundamental and second radial overtone modes. 
Furthermore, many linear combinations in the form $kf_0\pm f_2$, $k=0,1,2,\dots$ are present in the 
spectrum. As the period doubling effect has a time-dependent nature, we have 
checked whether the mode connected to $f_2$ is also temporally excited or not. We 
tested this possibility both with the analytical function method and the amplitude 
variation tool of {\sc Period04}. These two independent methods yielded very similar 
results and showed the amplitude of $f_2$ changing over the Blazhko cycle (see the 
plot below the light curve of V354\,Lyr in Fig.\,\ref{add}). This means that 
V354\,Lyr is a double mode pulsator with Blazhko effect, but not in a traditional 
sense.

\begin{figure}
\includegraphics[width=8cm]{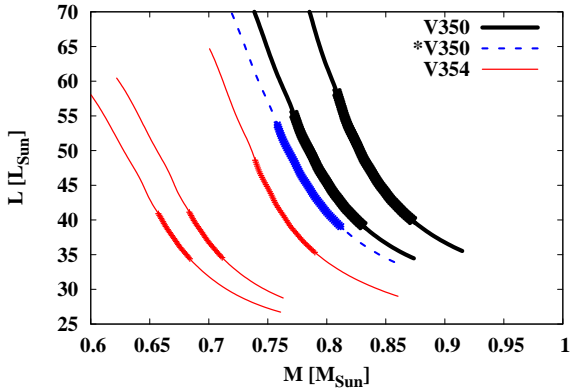}
\caption[]{
Linear pulsation models of V350\,Lyr and V354\,Lyr. 
The observed frequencies were fitted with the theoretical fundamental and second 
overtone modes. The lines represent the mass-luminosity values of the models which exactly
fit the observations. Thin (red) lines: V354\,Lyr periods with metallicities (left to right): 
Z=0.0001, 0.0003, 0.001; thick (black) lines: V350\,Lyr periods with
metallicities (left to right) Z=0.0001; 0.0003; dashed (blue) line: V350\,Lyr
periods with Z=0.0001 and the period ratio decreased by 0.001 
to show the effect of possible non-linear period shifts. The
thicker parts of the curves indicates the parameter range where the
second overtone is linearly unstable.
} \label{theo}
\end{figure}

The observed frequencies $f_0$ and $f_2$ can be easily matched by
the fundamental and second overtone modes in linear pulsation models
(Fig.\,\ref{theo}). We used the Florida--Budapest code \citep{Ye98, Kol02} 
to fit the observed frequencies by the theoretical fundamental and second overtones. 
The model frequencies depend on four parameters, the mass, luminosity,
effective temperature and chemical composition. Fixing two frequencies
reduces the unknown parameters by two. Then for given chemical
composition and effective temperature the mass and luminosity of the
matching star can be determined by  numerical model calculations. The
resulting mass-luminosity pairs are displayed on the Fig.\,\ref{theo}.
In the figure the linear instability region of the second
overtone is also displayed. We have to note, however, that the width
of these regions strongly depends on turbulent/convective parameters,
and in addition, the linear instability alone  is not a sufficient
condition for double mode pulsation. The coexistence of the
fundamental and second overtone modes in non-linear pulsation is a new
enigma that further theoretical studies should follow.

The frequencies $f^\prime$ ($2.0810\pm 0.0002$\,d$^{-1}$; $\sigma_{\mathrm s}=28.1$) 
$f^{\prime\prime}$ ($2.4407\pm 0.0003$\,d$^{-1}$; $\sigma_{\mathrm s}=11.6$) and 
$f^{\prime\prime\prime}$  ($2.6513\pm 0.0003$\,d$^{-1}$;  
$\sigma_{\mathrm s}=10.8$, insert in Fig.\,\ref{add}) are more difficult to identify. 
Their linear combinations with $f_0$ and its harmonics also appear 
in the spectrum and ratios are: $f_0/f^\prime 
=0.855$, $f_0/f^{\prime\prime}=0.729$, $f_0/f^{\prime\prime\prime}=0.672$. The 
last two ratios permit an explanation as a first overtone ($f^{\prime\prime}=f_1$) 
and a period-doubled ($f^{\prime\prime\prime}=3f_0/2$) frequency, respectively. 
However, $f^\prime$ does not seem to be fit this picture. This frequency might be 
connected to a non-radial mode.

The star {\bf V2178\,Cyg} (KIC\,3864443) shows a similar, but simpler and a 
bit less significant, additional peaks than V354\,Lyr. 
In this case the highest peak can be found 
again at the second overtone frequency 
$f_2$ ($3.5089\pm 0.0002$\,d$^{-1}$; $\sigma_{\mathrm s}=15.8$). One further peak 
at $f'=3.0585\pm 0.0003$\,d$^{-1}$ ($\sigma_{\mathrm s}=9.8$) 
was detected that might be a sign of a marginal period doubling effect ($3f_0/2$).

The other star where we found highly significant additional frequencies is 
{\bf V360\,Lyr} (KIC\,9697825). The dominating extra peak in its spectrum is located at 
$f_1=2.4875\pm 0.0001$\,d$^{-1}$ with $\sigma_{\mathrm s}=51.9$ 
(bottom right in Fig.\,\ref{add}). The ratio of this 
frequency to the fundamental mode frequency 
($f_0=1.79344\pm 0.00003$\,d$^{-1}$) is 0.721. This value is 
a bit smaller than the generally accepted ratio of fundamental to first radial 
overtone frequencies ($f_1/f_0=0.745$), but some model calculations with higher 
metallicity (see Fig.\,8 in \citealt{Cha10}) allow double mode pulsation with this 
ratio as well. We also detect a low-amplitude, but significant 
($\sigma_{\mathrm s}=18.9$), peak at the frequency of $f^\prime=2.6395\pm 0.0002$\,d$^{-1}$. 
Its ratio $f_0/f^\prime =0.679$ is almost 
the same as was found for V1127\,Aql, a Blazhko RR\,Lyrae star observed with CoRoT 
and explained there by a non-radial pulsation mode \citep{Cha10} .

{\bf V445\,Lyr} (KIC\,6186029) has a very complex structure 
of frequencies at the low-amplitude level 
(see right panel in Fig.\,\ref{double}). Here, the three types of frequency 
patterns (connected to the first and second overtone modes and period doubling 
effect) show similarly significant peaks. In decreasing order of amplitude: 
$3f_0/2=2.9231\pm 0.0002$\,d$^{-1}$ ($\sigma_{\mathrm s}=24.5$), 
$f_1=2.7725\pm 0.0002$\,d$^{-1}$ ($\sigma_{\mathrm s}=21.1$) and 
$f_2=3.331549\pm 0.0002$\,d$^{-1}$ ($\sigma_{\mathrm s}=13.5$).

If the global physical parameters are varying over the Blazhko cycle as 
demonstrated by \cite{JS09, JH09} excitation of a radial (or maybe a non-radial) 
mode temporally, (when the physical conditions of excitation are realized) would 
be a natural explanation for these patterns.

\begin{figure}
\includegraphics[width=8cm]{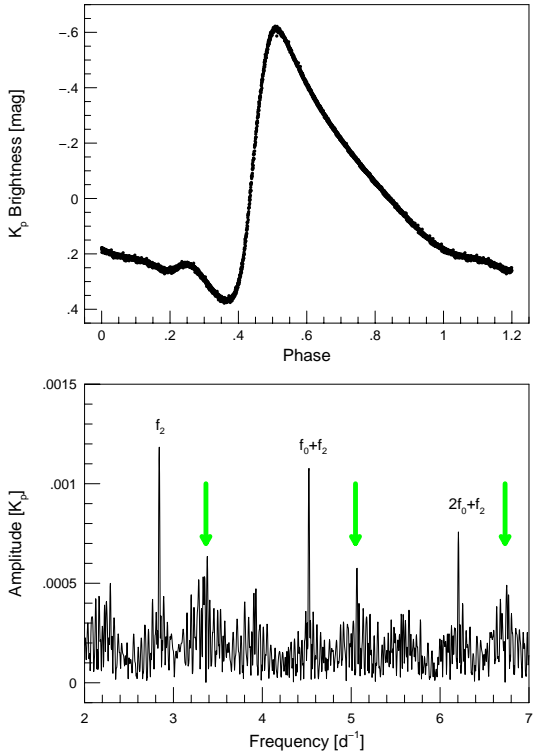}
\caption[]{
{\bf Top:} Phase diagram of V350\,Lyr (KIC\,9508655). The {\it Kepler} 
light curve is folded by the main period of $P_0=0.59424$~d. 
{\bf Bottom:} Fourier amplitude spectrum of V350\,Lyr around the
highest peaks after the data were prewhitened with the
main frequency $f_0$ and its harmonics. Due to small instrumental trends
some residuals are seen at the location of removed harmonics (shown by green arrows).
} \label{v350_fig}
\end{figure}

The frequency spectra of non-Blazhko stars were checked to see whether any of them 
possesses additional frequencies. This search resulted in the discovery the double 
mode nature of {\bf V350\,Lyr} (KIC\,9508655). 
Its Fourier spectrum (see Fig.\,\ref{v350_fig}) contains a significant 
($\sigma_{\mathrm s}=11.5$) peak at $f_2=2.8402\pm 0.0003$\,d$^{-1}$ with the amplitude 
of $A(f_2)=0.001$\,mag. The ratio of 
this frequency to the fundamental mode frequency ($1.682814\pm 0.00003$\,d$^{-1}$) 
is 0.592. That is, V350\,Lyr is the first example of a non-Blazhko double mode 
RR\,Lyrae star, where the fundamental and second overtone modes are excited. 
The extremely high amplitude ratio, $A(f_0)/A(f_2)=316$, points out why we have not 
been able to find such stars from the ground.

The residual peaks around the main frequency and its harmonics 
raise the possibility of the presence of the Blazhko effect close to the detection limit.
However, neither the $A_1(t)$ function nor the amplitude of the overtone frequency  
show clear time variation. In the future, calibrated {\it Kepler} data gathered
over a longer time span may clarify the nature of this interesting object.

In the case of V350\,Lyr, the match of the observed frequencies to the theoretical
model is a tougher problem than the case of V354\,Lyr. 
High mass or luminosity is required to fit the empirical 
data even at low metallicity (Fig.\,\ref{theo}).
 However, a small shift in the period ratio results in a
better agreement with the canonical mass and luminosity values. 0.001
difference in the observed period ratio and the linear model values
can be accounted to observational errors and non-linear effects of
stellar pulsations.

\section{Summary}

In this paper we have outlined some results obtained for RR\,Lyrae stars based on 
the first 138-d long data sets of the {\it Kepler Mission}: 
\begin{itemize}
\item
We have determined the 
main pulsation periods and amplitudes for all stars in this sample. 
These parameters were previously unknown or wrong for 9 stars.
\item
We have found 14 certain Blazhko stars among the 29 stars 
of the observed sample (48\%), 
and we could not find any deviations from monoperiodicity for 15 stars (52\%). 
Statistical distribution of the measured modulation amplitudes 
proved to be highly dependent on the used size of bins. Our sample  
supports the uniform distribution.
\item
The possibility of multiple modulations of an RR\,Lyrae star is well known,
such as the 4-year cycle of RR~Lyr itself. 
Further secondary modulation cycles were also reported in the literature.
We found a long term variation of the Blazhko cycle in V445\,Lyr. The 
forthcoming long time base of the {\it Kepler} data will allow us to investigate 
this phenomenon in detail.
\item
The {\it Kepler} data made it possible to find the Blazhko modulation of 
KIC\,1125706 with an amplitude as small as 0.03 mag. This is by far the smallest 
modulation amplitude ever detected for a Blazhko star. The same is true for phase 
modulations: we found small but clear phase variations for stars V2178\,Cyg, 
V783\,Cyg, V349\,Cyg, and KIC\,1125706. For these cases the $\delta P_0 < 1.5$\,min 
period variation again have the smallest known values.
\item
The sensitivity both in amplitude and phase made it possible to detect phase 
variation for all Blazhko stars. Although our sample is small, it is notable that 
all of our Blazhko stars are modulated both in their amplitude and phase 
simultaneously. The relative strength of the two types of modulation varies from 
star to star, but always has a common period. Therefore, a plausible explanation 
for the Blazhko effect must account for both amplitude and phase variations.
\item
We found additional frequencies, beyond the main frequency, its harmonics and 
expected modulation components, in the Fourier spectra of 7 Blazhko type stars. 
These additional frequencies concentrate around $f_0/2, 3f_0/2,\dots$ for three 
stars, while they appear around the first and second overtone frequency and its 
linear combination with the harmonics of the main frequency in the the case of 
V354\,Lyr, V2178\,Cyg and V360\,Lyr, respectively. A special case is V445\,Lyr 
where all of the above-mentioned types of frequencies are present.

If the basic physical parameters, such as mean radius, luminosity, effective 
temperature, are varying over the Blazhko cycle, a radial (or maybe non-radial) 
mode could be excited temporally. This would be a natural explanation for these 
transients.
\item
As a by-product of our frequency search for additional frequencies, we may have found the 
first double mode RR\,Lyrae star, V350\,Lyr, which pulsates in its fundamental and 
second overtone mode simultaneously.
\end{itemize}
Comprehensive and more detailed study of the increasing {\it Kepler} 
RR\,Lyrae data sets are in progress and will be discussed in the near
future.

\section*{Acknowledgments}

Funding for this Discovery mission is provided by NASA's Science Mission 
Directorate. This project has been supported by the National Office for Reseach and
Technology through the Hungarian Space Office Grant No.~URK09350 and 
the `Lend\"ulet' program of the Hungarian Academy of Sciences. 
KK acknowledges the support of Austrian FWF projects T359 and 
P19962. The authors gratefully acknowledge the entire {\it Kepler} team, whose 
outstanding efforts have made these results possible.

\bsp

\label{lastpage}

\end{document}